\documentclass[aps,prd,twocolumn,groupedaddress,showpacs,showkeys]{revtex4-1}
\usepackage{graphicx}

\normalbaselines \topmargin -0.5 truein  \oddsidemargin 0.30
truein \evensidemargin 0.30 truein 
scaled \magstep2  

\begin{document}

\title{ARCHIMEDEAN-TYPE FORCE IN A COSMIC DARK FLUID: \\
III. BIG RIP, LITTLE RIP AND CYCLIC SOLUTIONS}

\author{Alexander B. Balakin\footnote{e-mail: Alexander.Balakin@ksu.ru} and Vladimir V.
Bochkarev\footnote{e-mail: Vladimir.Bochkarev@ksu.ru}}
\affiliation{Kazan Federal University, Kremlevskaya str.,
18, 420008, Kazan,  Russia}

\date{\today}

\begin{abstract}
We analyze late-time evolution of the Universe in the framework of the self-consistent model, in which the dark matter is influenced by the
Archimedean-type force proportional to the four-gradient of the dark energy pressure. The dark energy is considered as a fluid with the
equation of state of the relaxation type, which takes into account a retardation of the dark energy response to the Universe accelerated expansion.
The dark matter is guided by the Archimedean-type force, which redistributes the total energy of the dark fluid between two its constituents, dark energy and dark
matter, in the course of the Universe accelerated expansion. We focus on the constraints for the dark energy relaxation time parameter, for the dark energy equation of state parameter, and for the Archimedean-type coupling constants, which guarantee the Big Rip avoidance. In particular, we show that the Archimedean-type coupling protects the Universe from the Big Rip scenario with
asymptotically infinite negative dark energy pressure, and that the Little Rip is the fate of the Universe with the Archimedean-type interaction inside the dark fluid.
\end{abstract}

\pacs{04.20.-q, 04.40.-b, 04.40.Nr}
\keywords{Dark matter, dark energy, Archimedean-type interaction, accelerated expansion}

\maketitle

\section{Introduction}

The present time accelerated expansion of the Universe discovered in the observations of the Supernovae Ia \cite{1,2,3}
revived the discussion about the fate of our Universe. A psychologically jeopardy variant of the Universe future
is that the permanently increasing scale factor $a(t)$ and the Hubble function $H(t)$ can reach infinite values during a finite time interval.
The life-time of such Universe is finite, and the catastrophe of this type produces cosmic inertial (tidal) forces, which  destroy the bounds in all physical systems.
Various aspects of the singular scenaria of the future stage of the Universe evolution were discussed during last two decades (see, e.g., the review \cite{4}).
The specific term Big Rip (or Doomsday) entered the scientific lexicon after publication of the paper \cite{BR1}; nowadays this term indicates a new trend in theoretical cosmology
(see, e.g., \cite{BR5}-\cite{BR11}).
It seems to be uncomfortable for physicist to think that the Big Rip
is the fate of our Universe, probably, it is a reason that many authors consider the models, in which the Big Rip can be {\it avoided}. In particular, the Big Rip can be avoided in the $F(R)$, $f(T)$-gravity models and their modifications, in the models with Chaplygin gas, in the model for the dark energy with various effective time-dependent equations of state (see, e.g., \cite{avo1}-\cite{avo12}).
As a more optimistic variant of the Universe behavior one could consider the case, when the Universe life-time is infinite, and the Hubble function tends asymptotically to a constant, $H \to H_{\infty}$. The asymptotic regime of this type appears, in particular,
in the $\Lambda$CDM model, which converts at $t \to \infty$ into the de Sitter model with $H_{\infty}{=}\sqrt{\frac{\Lambda}{3}}$ ($\Lambda$ is the cosmological constant).
More general case, when $H(t) \to H_{\infty}{=}const$, but $H_{\infty}$ is not necessary equal to $\sqrt{\frac{\Lambda}{3}}$,  is indicated as Pseudo Rip in \cite{Pseudo}.
Various intermediate scenaria with the infinite Universe's life-time, in which the scale factor, the Hubble function and (probably) its time derivative tend to infinity, belong to the class indicated by the term Little Rip (see, e.g., \cite{LR1}-\cite{LR7} for references).

\subsection{On the classification of the models of the late-time Universe behavior}

In order to classify the models of three types mentioned above in more detail, we use the terminology (see, e.g., \cite{BR7}) based on the asymptotic properties of the
scale factor $a(t)$, Hubble function $H(t){=}\frac{\dot{a}}{a}$, and its time derivative $\dot{H}$, which are the basic quantities in the isotropic spatially homogeneous cosmological models of the Friedmann-Lema\^itre-Robertson-Walker (FLRW)-type with the metric
\begin{equation}
ds^2 = dt^2 {-} a^2(t)[(dx^1)^2 {+} (dx^2)^2 {+} (dx^3)^2] \,. \label{metric}
\end{equation}
(Here and below we use the units with $c{=}1$).
In fact this classification is originated from the idea that two bonded particles
are influenced by the inertial force proportional to the quantity
\begin{equation}
\frac{\ddot{a}}{a} = \dot{H}{+}H^2 = - H^2 q(t) \,, \label{devi0}
\end{equation}
where ${-}q$ is the acceleration parameter.
The internal structure of the physical system is assumed to be destroyed, when this inertial force exceeds the internal (e.g., intermolecular) forces.
We think that such force can also be indicated as a {\it tidal} force, if we take into account the following motives.
Let us start with the well-known equation of the world-line deviation (see, e.g., \cite{MTW})
\begin{equation}
\frac{D^2 n^i}{D\tau^2} = - R^i_{\ klm} U^k U^m n^l + {\cal F}^i
\,, \label{devi1}
\end{equation}
where $U^k{=}\frac{dx^k}{d\tau}$ is the velocity four-vector, $n^i{=}\frac{dx^i}{d\lambda}$ is the deviation four-vector, $\tau$ is the world-line parameter and $\lambda$ is the parameter describing the world-line family. The second term in the right-hand side of (\ref{devi1}) describes the contribution of the force of the non-gravitational origin (its structure for the electromagnetic interactions is described, e.g., in \cite{Devi}). The first term in the right-hand side of (\ref{devi1}) is interpreted as the tidal (curvature induced) force proportional to the convolution of the Riemann tensor $R^i_{\ klm}$ with the deviation four-vector.
Taking into account that in the FLRW model the velocity four-vector is of the form $U^i{=}\delta^i_0$, we can choose arbitrary direction, say $0x^1$, and calculate the corresponding tidal force component $\Re^1 {=}{-} R^1_{\ 010} n^1$. The result is $\Re^1 {=}\frac{\ddot{a}}{a} n^1$, thus the interpretation in terms of tidal force is equivalent to the one in terms of inertial force.
This force tends to infinity, when $\dot{H}$ or $H$ (or both of them) tend to infinity.
The standard classification uses the following terminology:

\vspace{2mm}
\noindent
1. Power-law inflation, if $H {=} \frac{h}{t}$, $H(t \to \infty) \to 0$;

\noindent
2. $\Lambda$CDM model $H {=} H_{\Lambda} \equiv \sqrt{\frac{\Lambda}{3}}$, where $\Lambda$ is the cosmological constant;

\noindent
3. Pseudo Rip, if $H(t \to \infty) \to H_{\infty} < \infty$;

\noindent
4. Little Rip, when $H(t \to \infty) \to \infty$, e.g., as $ h_{*} e^{\xi t}$;

\noindent
5. The future singularities at finite time $t_s$, when $H(t \to t_s) \to \infty$.

The fifth subclass can be specify, for instance, as follows (see \cite{BR7}):

\noindent
5.1. Big Rip (Type I), if  $a \to \infty $, e.g., as $a \to \frac{1}{(t-t_s)^{\nu}}$;

\noindent
5.2. Sudden singularity  (Type II), if the pressure $\Pi$ tends to infinity, but $a(t)$ and energy density $\rho$ remain finite;

\noindent
5.3. Singularity of the Type III, if  $\Pi \to \infty$ and $\rho \to \infty$, but $a(t)$ is finite;

\noindent
5.4. Singularity of the Type IV, if only $\dot{H} \to \infty$.

\subsection{The goal of this paper}

This paper is considered to be the third (final) one describing the fundamentals of the model of the Archimedean-type interaction between the dark energy (DE) and dark matter (DM).
In \cite{Arch1} we established the self-consistent model
in which the dark matter is influenced by the force proportional to the four-gradient of the dark energy pressure. In \cite{Arch2} we classified the corresponding scenaria of the Universe evolution with respect to a number of the transition points, in which the acceleration parameter ${-}q(t)$ takes zero values, and thus the epochs of decelerated expansion is replaced by the epoch of the accelerated expansion. In \cite{Arch3} we considered the application of the model to the problem of light propagation with non-minimal coupling in a cosmic dark fluid with an Archimedean-type interaction between the DE and DM, and described the so-called unlighted cosmological epochs, for which the effective refraction index $n(t)$, the phase and group velocities, $V_{\rm ph}(t)$ and $V_{\rm gr}(t)$, respectively, were imaginary functions of time.

In \cite{Arch1,Arch2,Arch3} we focused on the solutions with asymptotically {\it finite}
Hubble function and finite acceleration parameter, and claimed that the case of {\it infinite} $H$ and $q$ is the topic of special discussion. Now we systematically consider the last case in terms of Big Rip and Little Rip solutions. Why this consideration might be interesting for readers? There are at least three motives to complete our investigation.
The first motif is the following. The dark energy is considered in \cite{Arch1,Arch2} as a fluid with the simplest rheological property: the equation of state is of the relaxation type, i.e., it
takes into account a retardation of the dark energy response, and includes an extra parameter, $\xi$. The first question arises: is it possible to avoid the Big Rip by the appropriate choice of the relaxation parameter $\xi$? The second motif is connected with the effective constant of the Archimedean-type coupling, $\nu_{*}$. When this coupling is absent, the energy density and pressure of the dark matter decrease in the course of the Universe expansion, so that the dark matter does not play any role in the asymptotic regime. However, when the pressure of the dark energy, being negative, becomes infinite in the course of accelerated expansion of the Universe, the Archimedean-type coupling leads to an effective heating of the dark matter, so that the DE and DM contributions happen to be of the same order. The second question arises: for what values of the coupling parameter $\nu_{*}$ the Archimedean-type interaction is able to protect the Universe from the Big Rip singularity?
The third motif is that the pressure of the dark energy, $\Pi$, is the key element in our model, which guides the behavior of both state functions of the dark matter, the energy-density $E$ and pressure $P$ (via the Archimedean-type force), the energy-density $\rho$ of the dark energy (via the equation of state), thus predetermining the behavior of the Hubble function $H$ and its derivative $\dot{H}$ (via the Einstein equations). This means that in the model under consideration one can classify the asymptotic types of the Universe behavior using only one element, just the function $\Pi(t)$. This function, $\Pi(t)$, satisfies the key equation, which is the nonlinear differential equation of the second order and includes six guiding parameters: $\xi$, $\sigma$ and $\rho_0$ coming from the three-parameter equation of state for the dark energy, the effective coupling constant of the Archimedean-type interaction $\nu_{*}$, the starting energy, $E_{*}$, and temperature, $T_{*}$, of the dark matter. Thus, (keeping in mind the initial data $\Pi(t_0)$ and $\dot{\Pi}(t_0)$) we have a possibility to classify all the types of asymptotic behavior of the model using eight parameters. The question arises: in what domains of the 8-dimensional effective space of the guiding parameters the Big Rip is reliably avoided?

In order to answer these questions we organized this paper as follows. In Section II we remind briefly the master equations for the DE and DM evolution in the framework of the Archimedean-type model (Subsections IIA-IIC). In Subsection IID we describe our classification scheme, which is based on eight parameters and includes three bifurcation points: $\nu_{*}{=}0$ ($\nu_{*}$ is the Archimedean-type coupling constant); $\xi{=}0$ ($\xi$ is the DE relaxation parameter); $\sigma{=}{-}1$, the phantom-crossing point.
In Section III we focus on the analysis of the models indicated as models with asymptotic DE domination, for which the contribution of the DM to the Universe expansion is asymptotically negligible. In Subsection IIIA we describe the Big Rip, Pseudo Rip, Cyclic solutions with $\nu_{*}{=}0$ and $\xi{=}0$ in terms of two guiding parameters $\sigma$, $\rho_0$, and the initial value of the DE pressure.
In Subsection IIIB we focus on the models with $\nu_{*}{=}0$ and $\xi \neq 0$, thus analyzing the role of the DE relaxation parameter in various scenaria of the Big Rip and Little Rip formation (in terms of three guiding parameters $\xi$, $\sigma$, $\rho_0$, and two initial data for the DE).
In Section IV we consider the models with $\nu_{*}\neq 0$ and $\xi{=}0$ (in terms of three guiding parameters $\nu_{*}$, $\sigma$, $\rho_0$ and the initial value of the DE pressure), and show that the Archimedean-type coupling of the DM with the barotropic DE avoids the Big Rip regimes in the Universe expansion. In Section V we discuss the general case $\nu_{*}\neq 0$ and $\xi \neq 0$ qualitatively and numerically, and show that the Archimedean-type coupling of the DM with the DE of the rheological type converts the Big Rip regimes into the Little Rip, Pseudo Rip and Cyclic regimes of the Universe expansion.

\section{Master equations}

Let us remind briefly the master equations of the model of the Archimedean-type coupling between the dark energy and dark matter. These equations are derived in \cite{Arch1} and now we use them to analyze the late-time behavior of the Universe in the framework of this model.

\subsection{Equations for gravity field}

The FLRW-type isotropic spatially homogeneous cosmological model with two interacting constituents: the dark energy and dark matter are known to be described by two nontrivial Einstein equations
\begin{equation}
\dot{H} = - 4\pi G[ \rho + E + \Pi + P]
\,,  \label{EinREDU10}
\end{equation}
\begin{equation}
H^2 = \frac{8\pi G}{3}(\rho + E) \,. \label{EinREDU1}
\end{equation}
Here the dot denotes the derivative with respect to time,
the functions  $\rho(t)$ and $\Pi(t)$ are the energy-density and pressure of the dark energy, respectively,
$E(t)$ and $P(t)$ describe the corresponding state functions of the dark matter.
The cosmological constant $\Lambda$ is considered to be incorporated into the DE energy density $\rho$ and DE
pressure $\Pi$. As usual, $H(t) \equiv \frac{\dot{a}}{a}$ is the Hubble
function. The conservation law as the compatibility equation for the set (\ref{EinREDU10})-(\ref{EinREDU1}) has the form
\begin{equation}
\dot{\rho} + \dot{E} + 3H (\rho + E + \Pi + P) =0 \,.
\label{balance}
\end{equation}

\subsection{Dark matter description}

The energy density and pressure of the one-component dark matter can be effectively presented, respectively, by the integrals
\begin{equation}
E(x) = \frac{E_{*}}{x^3} \int_0^{\infty}
q^2 dq \sqrt{1{+}q^2 F(x)} \ e^{{-}\lambda_{*}
\sqrt{1{+}q^2}}\,, \label{e(x)}
\end{equation}
\begin{equation}
P(x) = \frac{E_{*}}{3x^3} \int_0^{\infty}
\frac{F(x) q^4 dq}{\sqrt{1{+}q^2 F(x)}} \
e^{{-}\lambda_{*} \sqrt{1{+}q^2}}\,.  \label{p(x)}
\end{equation}
The auxiliary function $F(x)$ of the dimensionless variable $x\equiv \frac{a(t)}{a(t_0)}$
\begin{equation}
F(x) = \frac{1}{x^2} \exp{\{2\nu_{*}
[\Pi(1){-}\Pi(x)]\}} \,, \label{FF}
\end{equation}
and convenient parameters
\begin{equation}
E_{*} \equiv \frac{N_{*} m_{*}
\lambda_{*}}{K_2(\lambda_{*})}\,,  \quad
\lambda_{*} \equiv \frac{m_{*}}{k_{({\rm
B})}T_{*}} \,, \label{E}
\end{equation}
contain the effective Archimedean-type coupling constant $\nu_{*}$, an
effective number density $N_{*}$, the mass $m_{*}$ and temperature $T_{*}$ of the leading sort of the
dark matter particles. The term
\begin{equation}
K_{s}(\lambda_{*}) \equiv \int_0^{\infty} dz \cosh{s
z} \cdot \exp{[-\lambda_{*} \cosh z]}   \,, \label{McD}
\end{equation}
is the modified Bessel function, $k_{({\rm B})}$ is the Boltzmann constant.

\subsection{Dark energy dynamics}

To describe the dark energy fluid we use the linear three-parameter equation of state
\begin{equation}
\rho(t) = \rho_0 + \sigma \Pi + \frac{\xi}{H(t)} \dot{\Pi} \,.
\label{simplest0}
\end{equation}
When $\sigma {=}0$ and $\xi{=}0$, Eq. (\ref{simplest0}) introduces the
model in which the dark energy relates to the cosmological constant
$\Lambda$. When $\rho_0{=}0$ and $\xi{=}0$, Eq. (\ref{simplest0}) gives the well-known linear relation $\Pi {=} w \rho$
with $w \equiv \frac{1}{\sigma}$.
The retardation of response in the dark energy evolution is taken into account by inserting the term containing the first derivative of the
pressure $\dot{\Pi}$. An equivalent scheme is widely used in the extended thermodynamics and rheology (see,
e.g., \cite{REO1}), in which the extended constitutive equation for the
thermodynamically coupled variables ${\bf X}$ and ${\bf Y}$ has the form
\begin{equation}
\tau \dot{{\bf X}} + {\bf X} = w {\bf Y} \,.
\label{ES5}
\end{equation}
Here $\tau$ is a relaxation time, a new coupling parameter of the model.
In the cosmological context $\tau$ is generally the function of time, $\tau(t)$.
We assume that $ \tau(t) {=} \frac{\xi}{\sigma H(t)}$, i.e., this relaxation time can be measured in natural
cosmological scale \cite{tau1}. Our ansatz here is that the
dimensionless parameter $\xi$ is constant.

When the quantities $\rho(t)$ and $\Pi(t)$ depend on time through
the scale factor $a(t)$ only, i.e., $\rho {=} \rho(a(t))$, $\Pi =
\Pi(a(t))$, the so-called $x$-representation  is convenient, which is based
on the following relations
\begin{equation}
\frac{d}{dt} = x H(x)
\frac{d}{dx} \,,  \label{diffa}
\end{equation}
\begin{equation}
t-t_0 = \int_1^{\frac{a(t)}{a(t_0)}}
\frac{dx}{x H(x)} \,. \label{diffa2}
\end{equation}
In these terms the balance equation (\ref{balance}) and the constitutive equation (\ref{simplest0}) give
the key equation for the pressure of the dark energy $\Pi(x)$
\begin{equation}
\xi x^2 \Pi^{\prime \prime}(x) {+} x \Pi^{\prime}(x) \left(4 \xi {+}
\sigma \right) {+} 3 (1{+}\sigma)\Pi {+}
3 \rho_0 {=} {\cal J}(x) \,.
\label{key1}
\end{equation}
The prime denotes the derivative with respect to $x$.
The source ${\cal J}(x)$ is defined as follows:
\begin{equation}
{\cal J}(x) \equiv {-} E_{*} \frac{\left[x^2
F(x)\right]^{\prime}}{2 x^4} \int_0^{\infty}\frac{q^4
dq e^{{-}\lambda_{*} \sqrt{1{+}q^2}}}{\sqrt{1{+}q^2 F(x)}} .
\label{key2}
\end{equation}
The quantity ${\cal J}(x)$ vanishes, when the Archimedean-type coupling parameter vanishes, i.e.,
$\nu_{*}=0$. We deal with the differential equation of the second-order linear in the derivatives but nonlinear in the unknown
function $\Pi(x)$.

\subsection{Classification scheme based on the analysis of the DE pressure $\Pi$ }

The function $\Pi$ plays the key role in the analysis of the whole model behavior. Indeed, when $\Pi(x)$ is found from
the key equation (\ref{key1}), we can reconstruct the DE energy-density $\rho$ using the formula
\begin{equation}
\rho(x) = \rho_0 + \sigma \Pi(x) + \xi x \Pi^{\prime}(x) \,,
\label{simplest1}
\end{equation}
then calculate the DM state functions $E(x)$ and $P(x)$ using (\ref{e(x)}) and (\ref{p(x)}), then find $H(x)$ from the
Einstein equation (\ref{EinREDU1}), and finally, reconstruct the scale factor as the function of time using (\ref{diffa2}).

We are interested to discuss the late-time period of evolution, i.e., the period, when $t \to t_{\infty}$. We consider both cases:
$t_{\infty} {=}t_s$ (future finite time singularity)  and
$t_{\infty} {=} \infty$. Clearly, there are three possible types of the asymptotic behavior of the dark fluid composed of the dark energy and dark matter.

\noindent
1. The first type refers to the asymptote $\Pi(t \to t_{\infty}) \to \Pi_{\infty}
{=} const$. This case includes the submodel with
$\Pi_{\infty}{=}0$). Since the Archimedean-type force is
proportional to the four-gradient of the dark energy pressure
$\Pi$, then asymptotically the dark matter decouples from the dark
energy, its energy-density $E$ decreases as $\frac{1}{a^3}$ and we obtain the case indicated as an asymptotic DE domination.

2. The second type of behavior refers to the case $\Pi(t \to t_{\infty}) \to
{+}\infty$. Now the DM becomes frozen, since the function $F(x)$ (\ref{FF})
tends to zero exponentially,  the DM decouples from the DE,
and again we deal with the model of the DE domination.

3. The third and the most interesting case, refers to the asymptotic
behavior $\Pi(t \to t_{\infty}) \to {-}\infty$, for which the function $F(x)$ tends to
infinity, the DM becomes effectively ultrarelativistic and thus
plays an active role in the energy redistribution process.

The model as a whole contains four effective guiding parameters ($\xi$, $\sigma$ and  $\rho_0$ describing the DE equation of state, and $\nu_{*}$,
the effective Archimedean-type coupling constant) and four initial parameters ($\Pi(t_0)$ and $\dot{\Pi}(t_0)$, initial data for the DE pressure,
and $E_{*}$, $T_{*}$ initial data for the DM). The classification scheme includes three bifurcation points: the first, $\nu_{*}{=}0$ indicates whether the Archimedean-type coupling is switched on or not; the second, $\xi{=}0$ indicates whether the DE possesses the simplest rheological property or not, the third,  $\sigma{=}{-}1$, relates to the phantom-crossing point.

\vspace{3mm}

\noindent
{\it Remark on the Big Rip symptom:}

\vspace{3mm}
\noindent
when the integral in the right-hand side of (\ref{diffa2}) converges on the upper limit,
the infinite value of the scale factor $a(t)$ can be reached at finite time $t_{({\rm s})}$, thus we deal with the Big Rip.

\section{Big Rip solutions in the models with asymptotic dark energy domination}

\subsection{Big Rip solutions at $\xi{=}0$ and $\nu_{*}{=}0$ }

This first model relates to the case, when the dark energy does not possess rheological properties, and there is no Archimedean-type coupling between DE and DM.
The contribution of the DM is asymptotically vanishing, and in this sense we deal with the case of the DE domination. The main results for this model are well-known and our goal is to recover them as the limiting case of our model in the corresponding terminology.

\subsubsection{Power-law expansion with $\rho_0{=}0$}

Let us recover, first, the well-known results presenting the Big Rip solutions, when $\xi{=}0$, $\rho_0{=}0$ and $\nu_{*}{=}0$.
The key equation (\ref{key1}) takes the form
\begin{equation}
\sigma  x \Pi^{\prime}(x) {+} 3 (1{+}\sigma)\Pi {=} 0 \,,
\label{key11}
\end{equation}
and the solution is
\begin{equation}
\Pi(x) = \Pi(1) \ x^{-3\frac{(1{+}\sigma)}{\sigma}} \,.
\label{key111}
\end{equation}
It is convenient to introduce the new dimensionless parameter
\begin{equation}
\alpha \equiv -\frac{\sigma}{3(1{+}\sigma)} \,,
\label{key1119}
\end{equation}
since it plays an important role below.
For the DE energy-density $\rho(x)$ we immediately obtain
\begin{equation}
\rho(x) = \sigma \Pi(1) \ x^{\frac{1}{\alpha}} \,,
\label{key1118}
\end{equation}
so that
\begin{equation}
H^2(x) = \frac{1}{3}8\pi G \sigma \Pi(1) \ x^{\frac{1}{\alpha}}  \,.
\label{key121}
\end{equation}
Clearly, the real solution for the Hubble function exists, when the product $\sigma \Pi(1)$ is positive.
Using the remark on the Big Rip symptom one can conclude that the integral in (\ref{diffa2}) converges, when $\alpha$ is positive, i.e. ${-}1<\sigma<0$.
This simplest submodel allows us to verify this fact directly by computing the scale factor analytically.
Indeed, according to (\ref{diffa2}) the scale factor has the form
\begin{equation}
\frac{a(t)}{a(t_0)}{=} \left[1{-} \frac{1}{3\alpha} \sqrt{6 \pi G \sigma \Pi(1)}(t{-}t_0) \right]^{-2\alpha} \,.
\label{key141}
\end{equation}
When ${-}1<\sigma<0$ or in other words, when $w{=}\frac{1}{\sigma}<{-}1$, one obtains that the parameter $\alpha$ is positive. This means that
the scale factor tends to infinity at some finite time value $t{=}t_{*}$, given by
\begin{equation}
t_{*} = t_0 {+} \frac{3 \alpha}{\sqrt{6 \pi G \sigma \Pi(1)}} > t_0 \,.
\label{key17}
\end{equation}
When $\sigma<0$, the solution is real, when the initial value of the DE pressure is negative
$\Pi(1)<0$. Thus we deal with the well-known Big Rip solution, which is characterized by the following asymptotes at $t \to t_{*}$:  $\Pi \to {-}\infty$, $\rho \to {+} \infty$, $a \to \infty$, $H \to \infty$ and $\dot{H} \to \infty$. Let us add that when the parameter $\sigma$ is negative and belongs to
the interval ${-}1<\sigma<0$, the parameter $\alpha$ is
positive (see (\ref{key1119})). There are two special cases. When
$\sigma \to {-1}$, the parameter $\alpha$ becomes infinite,
$\alpha \to {+} \infty$; when $\sigma \to 0$, the parameter
$\alpha$ tends to zero also. Finally, when $\alpha$ is negative, i.e., $\sigma>0$ or $\sigma<{-}1$, the solution (\ref{key141}) describes power-law expansion of the Universe.
The Little Rip solutions do not appear in this submodel.

\subsubsection{Solutions with $\rho_0 \neq 0$}

When $\rho_0 \neq 0$, the solution of the key equation
\begin{equation}
\sigma  x \Pi^{\prime}(x) {+} 3 (1{+}\sigma)\Pi {+} 3 \rho_0 {=} 0
\label{key119}
\end{equation}
has the form
\begin{equation}
\Pi(x) {=} {-} \frac{\rho_0}{1{+}\sigma} {+}  x^{\frac{1}{\alpha}} \left[\Pi(1){+} \frac{\rho_0}{1{+}\sigma} \right] \,.
\label{key115}
\end{equation}
The square of the Hubble function reads now
\begin{equation}
H^2(x) {=} \frac{8\pi G \rho_0}{3(1{+}\sigma)} \left[1 {+}  {\cal A} x^{\frac{1}{\alpha}} \right] \,,
\label{key12}
\end{equation}
where
\begin{equation}
{\cal A} \equiv \sigma \left[1 {+} \frac{(1{+}\sigma)\Pi(1)}{\rho_0} \right] \,.
\label{key13}
\end{equation}
The signs of the parameters ${\cal A}$, $\rho_0$, $\alpha$ and $\sigma$ provide important details of the classification. In order to classify the models we, first, distinguish  three (principal) cases and then show that all other cases can be reduced to these three ones.

\vspace{3mm}
\noindent
{\it (1) The model with ${\cal A}>0$, ${-}1<\sigma<0$, $\rho_0 >0$.}

\noindent
Let us consider, first, the case when ${-}1<\sigma<0$ and ${\cal A}>0$. The corresponding initial value $\Pi(1)$ should be negative, and the inequality
$\Pi(1)<{-}\frac{\rho_0}{1{+}\sigma}$ has to be satisfied. The solution for the scale factor has the form
\begin{equation}
\frac{a(t)}{a(t_0)}{=} \left[\cosh{\mu(t{-}t_0)} {-} \sqrt{1{+{\cal A}}} \sinh{|\mu|(t{-}t_0)} \right]^{{-}2\alpha}
\,,
\label{key14}
\end{equation}
where
\begin{equation}
\mu \equiv \frac{1}{\sigma} \sqrt{6 \pi G \rho_0 (1{+}\sigma)} \,.
\label{key15}
\end{equation}
There exists a moment $t{=}t_{*}$, for which $a(t_{*}){=}\infty$. At $t \to t_{*}$ the plot of $a(t)$ has a vertical asymptote.
This critical moment of time, $t_{*}$, can be found from the equation
\begin{equation}
\tanh{|\mu|(t_{*}{-}t_0)} =  \frac{1}{\sqrt{1+ {\cal A}}} \,,
\label{key16}
\end{equation}
and has the form
\begin{equation}
t_{*} = t_0 + \frac{1}{2|\mu|} \log{\left(\frac{\sqrt{1{+} {\cal A}}{+}1}{\sqrt{1{+} {\cal A}}{-}1}\right)} > t_0 \,.
\label{key167}
\end{equation}
Again we deal with the Big Rip solution, characterized by the asymptotes $\Pi \to {-} \infty$, $\rho \to {+}\infty$, $a \to \infty$, $H \to \infty$, $\dot{H} \to \infty$, but now, in the case $\rho_0 \neq 0$, we can  indicate the Big Rip singularity as that of hyperbolic type in contrast to the one of the power-law type, presented in the previous subsection (with $\rho_0 {=}0$).

\vspace{3mm}
\noindent
{\it Remark 1:}

\noindent
Two submodels with negative $\alpha$, namely, ${\cal A}>0$, $\sigma >0$, $\rho_0 >0$ and
${\cal A}>0$, $\sigma<{-}1$, $\rho_0 <0$, give formally the same scale factor (\ref{key14}). However, the essential difference is that at $\alpha <0$ the Hubble function $H$ (\ref{key12}) tends
to the constant $H \to H_{\infty} {=} \sqrt{\frac{8\pi G \rho_0}{3(1{+}\sigma)}}$, and we deal with the Pseudo Rip instead of the Big Rip.

\vspace{5mm}
\noindent
{\it (2) The model with
${\cal A}<0$, ${-}1<\sigma<0$, $\rho_0 >0$.}

\noindent
When ${-}1<\sigma<0$ and ${\cal A}< 0$, one obtains from (\ref{key12}), that there is a critical value $x{=}x_{*}$
\begin{equation}
x_{*} \equiv |{\cal A}|^{- \alpha} \,,
\label{key27}
\end{equation}
for which the function $H^2(x)$
\begin{equation}
\quad H^2(x) {=} \frac{8 \pi G \rho_0}{3(1{+}\sigma)}\left[1 {-} \left(\frac{x}{x_{*}} \right)^{\frac{1}{\alpha}} \right]
\label{key28}
\end{equation}
takes zero value. Let us assume that $x_{*}>1$, i.e., according to (\ref{key27}) $|{\cal A}|<1$ and ${-}1<{\cal A}<0$. It is possible when
${-}\frac{\rho_0}{(1{+}\sigma)}<\Pi(1)<\frac{\rho_0}{|\sigma|}$.
Since the model solutions for $H(x)$ obtained from (\ref{key28}) can not be prolonged for $x>x_{*}$, $x$ starts to decrease after the moment
$t{=}t_{({\rm max})}$. The corresponding solution for the scale factor
has now the form
\begin{equation}
\frac{a(t)}{a(t_0)}{=} \left[\frac{\cosh{\mu(t_0{-}t_{({\rm max})})}}{\cosh{\mu(t{-}t_{({\rm max})})}} \right]^{2\alpha}
\,,
\label{key24}
\end{equation}
where
\begin{equation}
t_{({\rm max})} = t_0 + \frac{1}{2|\mu|} \log{\left(\frac{1{+}\sqrt{1{-}|{\cal A}|}}{\sqrt{|{\cal A}|}}\right)} > t_0 \,.
\label{key1267}
\end{equation}
Clearly, the plot of this function is symmetric in reference to the moment $t_{({\rm max})}$; the formula (\ref{key24}) gives the maximal value (\ref{key27})
$a(t_{*}){=}a(t_0) |{\cal A}|^{-\alpha}$ at this moment. At $t \to \infty$ the scale factor $a(t)$ tends to zero as
\begin{equation}
a(t) \propto \exp\{{-} H_{\infty} t \} \,, \quad H_{\infty} \equiv \sqrt{\frac{8\pi G \rho_0}{3(1{+}\sigma)}} \,.
\label{key0167}
\end{equation}
The Hubble function is described by the finite function
\begin{equation}
H(t)= {-}  H_{\infty} \tanh{\left[|\mu|(t{-}t_{({\rm max})})\right]} \,,
\label{key249}
\end{equation}
which changes the sign at $t{=}t_{({\rm max})}$. This means that the Universe expansion turns into a collapse at this moment.
The acceleration parameter
\begin{equation}
{-}q(t) {=} \frac{\ddot{a}}{a H^2} {=} 1 {-} \frac{1}{2\alpha}\sinh^{{-}2}{\mu(t{-}t_{({\rm max})})}
\label{key259}
\end{equation}
takes infinite value at $t{=}t_{({\rm max})}$, since the Hubble function in the denominator vanishes at this moment.
Clearly, the Universe passes two eras with $-q>0$ (accelerated expansion  and accelerated collapse), and two eras with $-q<0$ (decelerated expansion and decelerated collapse).
We deal with a closed Universe which does not admit the Big Rip behavior, despite ${-1}<\sigma<0$.

\vspace{3mm}
\noindent
{\it Remark 2:}

\noindent
Two submodels with negative $\alpha$, namely, ${\cal A}<0$, $\sigma >0$, $\rho_0 <0$ and
${\cal A}<0$, $\sigma<{-}1$, $\rho_0 >0$, give formally the same expression for the scale factor (\ref{key24}). Since now $\alpha$ is negative,  the Hubble function $H$ (\ref{key12}) tends
again to the constant $H \to H_{\infty} {=} \sqrt{\frac{8\pi G \rho_0}{3(1{+}\sigma)}}$, if we require $x_{*}<1$ to avoid the internal singular point. We deal now with the Pseudo Rip instead of Big Rip.

\vspace{5mm}
\noindent
{\it (3) The model with
${\cal A}<0$, ${-}1<\sigma<0$, $\rho_0 <0$.}

\noindent
Now the scale factor can be expressed in terms of trigonometric functions
\begin{equation}
\frac{a(t)}{a(t_0)}{=} \left[\cos{|\mu|(t{-}t_0)} {-} \sqrt{|{\cal A}|{-}1} \sin{|\mu|(t{-}t_0)} \right]^{-2\alpha}
\,,
\label{key34}
\end{equation}
the parameter ${\cal A}$ has to satisfy the inequality ${\cal A}<{-1}$, thus, the initial value $\Pi(1)$ is restricted by $\Pi(1)<{-}\frac{|\rho_0|}{|\sigma|}$.
Again, there exists a moment $t_{*}$
\begin{equation}
t_{*} = t_0 + \frac{1}{|\mu|} {\rm arctg}\left(\frac{1}{\sqrt{|{\cal A}|{-}1}}\right) > t_0 \,,
\label{key1367}
\end{equation}
for which the plot of the scale factor has the vertical asymptote typical for the Big Rip with $\Pi \to {-} \infty$, $\rho \to {+}\infty$, $a \to \infty$, $H \to \infty$, $\dot{H} \to \infty$.
It is convenient to indicate such singularity as the Big Rip singularity of the trigonometric type.

\vspace{3mm}
\noindent
{\it Remark 3:}

\noindent
Two submodels with negative $\alpha$, namely, ${\cal A}<0$, $\sigma >0$, $\rho_0 <0$ and
${\cal A}<0$, $\sigma<{-}1$, $\rho_0 >0$, give formally the same scale factor (\ref{key34}). Since now the parameter $\mu$ becomes imaginary (see (\ref{key15})), at $\alpha <0$ one should replace the trigonometric functions by the hyperbolic one. Again, asymptotically we deal  with the Pseudo Rip solution instead of Big Rip.

\subsection{Big Rip and Little Rip solutions at $\xi \neq 0$ and $\nu_{*}{=}0$}

When we consider this submodel we assume that the Archimedean-type coupling is switched of, but the dark energy possesses the simplest rheological properties.
Since $\nu_{*}{=}0$, the asymptotic regime again will be characterized by the dark energy domination. In order to analyze this model let us introduce the new
unknown function $Z(x)$ as follows:
\begin{equation}
\Pi(x) {=} {-} \frac{\rho_0}{1{+}\sigma} {+} Z(x)  \,, \label{spec17}
\end{equation}
and define the initial value $Z(1)$ as
\begin{equation}
Z(1) = \Pi(1) + \frac{\rho_0}{1{+}\sigma}  \,. \label{spec171}
\end{equation}
The function $Z(x)$ satisfies the Euler equation
\begin{equation}
\xi x^2 Z^{\prime \prime}(x) + x Z^{\prime}(x) \left(4\xi + \sigma \right) + 3 (1+\sigma) Z  = 0
\,.
\label{key179}
\end{equation}
The characteristic polynomial of this Euler equation has two roots
\begin{equation}
s_{1,2} = \frac{1}{2\xi} \left[- (\sigma + 3\xi) \pm \sqrt{(\sigma -
3\xi)^2 -12\xi} \right] \,, \label{roots17}
\end{equation}
which can be real or complex depending on the values of the parameters $\xi$
and $\sigma$. One can distinguish three subcases.

\subsubsection{Two different real roots [$(\sigma-3\xi)^2 > 12\xi$]}

\noindent
{\it (i)} $\xi>0$.

\noindent
When the discriminant in (\ref{roots17}) is positive, one obtains
$$
Z(x) {=} x^{{-} \gamma} \left\{
\frac{Z(1)}{2} \left(x^{\Gamma} {+}x^{{-}\Gamma} \right) {+}
\right.
$$
\begin{equation}
\left.
{+} \frac{\left[Z^{\prime}(1){+} \gamma Z(1) \right]}{2\Gamma}  \left(x^{\Gamma} {-} x^{{-}\Gamma} \right)  \right\} \,, \label{aperiod2}
\end{equation}
where
\begin{equation}
\gamma \equiv \frac{\sigma + 3 \xi}{2\xi} \,, \quad \Gamma \equiv \frac{1}{2\xi} \sqrt{(\sigma
-3\xi)^2 -12\xi} \,. \label{aperiod1}
\end{equation}
According to the remark about the Big Rip symptom we can conclude that the integral in (\ref{diffa2}) converges now, when  $\Gamma > \gamma$, thus this inequality guarantees
the Big Rip existence. In order to {\it illustrate} the existence of the Big Rip by exact solutions, let us assume that initial data satisfy the equality
\begin{equation}
Z^{\prime}(1) = Z(1)(\Gamma - \gamma) = (\Gamma - \gamma) \left[\Pi(1)+ \frac{\rho_0}{1{+}\sigma} \right] \,. \label{ap1}
\end{equation}
Then the solutions can be simplified as follows
\begin{equation}
Z(x) {=} Z(1) \ x^{\Gamma {-} \gamma} \,,
\label{0ap2}
\end{equation}
and $Z(x)$ becomes proportional to the required leading order term exactly. Now the DE state functions are of the form
\begin{equation}
\Pi(x){=} {-} \frac{\rho_0}{1{+}\sigma} {+} Z(1) \ x^{\Gamma {-} \gamma} \,,
\label{ap2}
\end{equation}
\begin{equation}
\rho(x) {=} \frac{\rho_0}{1{+}\sigma} + \Omega(\xi) Z(1)\ x^{\Gamma {-} \gamma} \,,
\label{ap3}
\end{equation}
where the new parameter
\begin{equation}
\Omega(\xi) \equiv \sigma {+} \xi (\Gamma {-} \gamma)
\label{0ap3}
\end{equation}
is introduced.
In the case of the DE domination the square of the Hubble function
\begin{equation}
H^2(x){=} \frac{8 \pi G \rho_0}{3(1{+}\sigma)} \left[1 {+} \frac{\Omega(\xi) (1{+}\sigma)}{\rho_0} \ Z(1) \ x^{\Gamma {-} \gamma} \right] \,,
\label{ap4}
\end{equation}
formally coincides with (\ref{key12}), if we use the substitutions
\begin{equation}
{\cal A} \to \frac{\Omega(\xi) (1{+}\sigma)}{\rho_0} \ Z(1) \,, \quad  \frac{1}{\alpha} \to \Gamma {-} \gamma \,.
\label{ap5}
\end{equation}
Based on the results of the previous section one can conclude that the Big Rip takes place, if $\Gamma > \gamma$.
There are two domains on the plane of the parameters $\sigma 0 \xi$, where this inequality holds at $\xi>0$: first,
${-}1 < \sigma <3\xi {-}2\sqrt{3\xi}$, $\xi < \frac{1}{3}$, second, $\sigma < {-}1$. The first domain is a natural extension of the interval
${-}1 < \sigma <0$, which is obtained if we put $\xi{=}0$ into the right-hand side of the inequality; let us remind that just for the case ${-}1 < \sigma <0$ the Big Rip appears, when $\xi{=}0$.
The second interval for the Big Rip existence, $\sigma < {-}1$ at $\xi>0$, has no analog at $\xi{=}0$.

The parameter $\rho_0$ distinguishes three situations admitting the Big Rip at $\xi >0$:

\vspace{2mm}
\noindent
{(1)} $\rho_0 {=}0$.

\noindent
In both domains the singularity of $a(t)$ is  of the power-law type.

\vspace{2mm}
\noindent
{(2)} $\rho_0 > 0$.

\noindent
In the first domain, where ${-}1 < \sigma <3\xi {-}2\sqrt{3\xi}$, $\xi < \frac{1}{3}$, the singularity is of the hyperbolic type.
In the second domain, where $\sigma < {-}1$, the singularity is of the trigonometric type.

\vspace{2mm}
\noindent
{(3)} $\rho_0 < 0$.

\noindent
In the first domain, where ${-}1 < \sigma <3\xi {-}2\sqrt{3\xi}$, $\xi < \frac{1}{3}$, the singularity is of the trigonometric type.
In the second domain, where $\sigma < {-}1$, the singularity is of the hyperbolic type.

More detailed constraints for the guiding parameters follow from the inequalities $\Omega(\xi) Z(1) >0$ or $\Omega(\xi) Z(1) <0$, respectively.

\vspace{3mm}
\noindent
{\it (ii)} $\xi<0$.

\noindent
When the parameter $\xi$ is negative, we use the following scheme of analysis.  We put
\begin{equation}
Z^{\prime}(1) = - Z(1)(\Gamma + \gamma)
\label{ap18}
\end{equation}
and obtain that
\begin{equation}
Z(x) {=} Z(1) \ x^{{-}(\Gamma {+} \gamma)} \,.
\label{1ap29}
\end{equation}
The DE state functions take now the form
\begin{equation}
\Pi(x){=} {-} \frac{\rho_0}{1{+}\sigma} {+} Z(1) \ x^{{-}(\Gamma {+} \gamma)} \,,
\label{7ap2}
\end{equation}
\begin{equation}
\rho(x) {=} \frac{\rho_0}{1{+}\sigma} + \Omega(\xi) Z(1)\ x^{{-}(\Gamma {+} \gamma)} \,,
\label{7ap3}
\end{equation}
and the square of the Hubble function reads
\begin{equation}
H^2(x){=} \frac{8 \pi G \rho_0}{3(1{+}\sigma)} \left[1 {+} \frac{\Omega(\xi) (1{+}\sigma)}{\rho_0} \ Z(1) \ x^{{-}(\Gamma {+} \gamma)} \right] \,.
\label{7ap4}
\end{equation}
Again, these functions can be obtained asymptotically at $x \to \infty$, since now $\Gamma<0$ and in this limit one obtains that $x^{\Gamma} \to 0$.
The sum $\Gamma {+} \gamma$ should be negative, and the product $\Omega(\xi) Z(1)$ should be positive, if we try to find the analogs of the Big Rip.
This is possible, when $\sigma > {-}1$. Clearly, we obtain the power-law singularity, when $\rho_0 {=}0$, the hyperbolic singularity, when $\rho_0 > 0$, and
the trigonometric singularity, when $\rho_0 > 0$.

\subsubsection{Double real roots [$(\sigma{-}3\xi)^2 {=} 12\xi$]}

The discriminant can take zero value, only when $\xi>0$. For the positive $\xi$ there are three interesting cases: $\xi<\frac{1}{3}$, $\xi>\frac{1}{3}$ and $\xi{=}\frac{1}{3}$.

\vspace{2mm}
\noindent
{\it (i)} $\xi<\frac{1}{3}$.

\noindent
The corresponding solution for $Z(x)$ is
\begin{equation}
Z(x) =  x^{-\gamma}\left\{Z(1){+} \log{x} \left[Z^{\prime}(1){+} \gamma Z(1)  \right]\right\}  \,, \label{aperiod22}
\end{equation}
where $\gamma {=} 3 {+} \sqrt{\frac{3}{\xi}}$, if $\sigma {=} 3\xi {+}
2\sqrt{3\xi}$, and $\gamma {=} 3 {-} \sqrt{\frac{3}{\xi}}$, if $\sigma {=} 3\xi {-} 2\sqrt{3\xi}$. According to the mentioned symptom, the Big Rip is possible, when $\gamma$ is negative; clearly,
it is possible only if we deal with the second solution, i.e., when $\xi<\frac{1}{3}$ and $\sigma {=} 3\xi {-} 2\sqrt{3\xi}$. These constraints require that ${-1}<\sigma<0$.
In order to illustrate the behavior of the scale factor, let us assume that, first, $\rho_0{=}0$, second, $Z^{\prime}(1){=}{-}Z(1)\frac{\sigma}{\xi}$ and, third, $Z(1)<0$. Then we obtain
\begin{equation}
\rho(x) = 3|Z(1)| \ x^{|\gamma|} \log{x} \,, \label{aperi1}
\end{equation}
\begin{equation}
|\gamma| {=} \sqrt{\frac{3}{\xi}}(\sqrt{3\xi}{-}1) \,. \label{aperi18}
\end{equation}
The square of the Hubble function reads
\begin{equation}
H^2(x) = 8\pi G |Z(1)| x^{|\gamma|} \log{x} \,,  \label{aperi2}
\end{equation}
and the integral in (\ref{diffa2})
\begin{equation}
\sqrt{4G |\gamma Z(1)|}(t{-}t_0) = {\rm erf}\left[\sqrt{\frac{1}{2}|\gamma| \ \log{x}} \right] \,,
\label{aperi3}
\end{equation}
happens to be reduced to the error-function
\begin{equation}
{\rm erf}[z] \equiv \frac{2}{\sqrt{\pi}} \int_0^{z} du \ e^{{-}u^2} \,. \label{aperi4}
\end{equation}
The error-function ${\rm erf}[z]$ takes finite value ${\rm erf}[\infty] {=} 1$, when the upper limit in the integral goes to the infinity, thus, the value $x \to \infty$ can be reached at the moment
$t_s {=}t_{0}{+}\frac{1}{\sqrt{4G |\gamma Z(1)|}}$.
The scale factor has the form
\begin{equation}
a(t){=}a(t_0) \exp\left\{
\frac{2}{|\gamma|}\left[{\rm erf^{{-}1}}[\sqrt{4G|\gamma Z(1)|}(t{-}t_0)] \right]^2 \right\} \,,
\label{aperi5}
\end{equation}
where the symbol ${\rm erf^{{-}1}}$ stands for the function inverse to the error-function.
Thus, when $\xi<\frac{1}{3}$ the Big Rip is possible, if $\sigma {=} 3\xi {-} 2\sqrt{3\xi}$.

\vspace{2mm}
\noindent
{\it (ii)} $\xi>\frac{1}{3}$.

\noindent
This case corresponds to the positive parameter $\gamma$. This means that $Z(x)$ (see (\ref{aperiod22})) asymptotically vanishes, the Hubble function tends to constant, and we deal with the Pseudo Rip.

\vspace{2mm}
\noindent
{\it (iii)} $\xi=\frac{1}{3}$.

\noindent
Double real roots appear now, when $(\sigma{-}1)^2 {=} 4$, i.e., when $\sigma{=}3$ or $\sigma{=}{-}1$. The first root, $\sigma{=}3$, corresponds to the solution
(\ref{aperiod22}) with $\gamma{=6}$, and we again deal with the Pseudo Rip. The second root, $\sigma{=}{-1}$, refers to the special case with diverging quantity $\frac{\rho_0}{1{+}\sigma}$, and we have to return to the key equation (\ref{key1}).
As we emphasized in \cite{Arch1}, in this special model the DE pressure and
the DE energy density contain the logarithmic terms squared. In particular, for the initial data, which satisfy the equality $\rho(1){+}\Pi(1){=}0$, these DE state functions are
\begin{equation}
\Pi(x) {=} \Pi(1) {-} 3 \rho_0 \log{x} {-} \frac{9}{2}\rho_0
\log^2{x} \,, \label{Sc31}
\end{equation}
\begin{equation}
\rho(x) = \rho(1) + \frac{9}{2}\rho_0 \log^2{x} \,. \label{Sc41}
\end{equation}
The scale factor has the form
\begin{equation}
\frac{a(t)}{a(t_0)} {=} \exp\left\{
\sqrt{\frac{2\rho(1)}{9\rho_0}} \sinh{\left[\sqrt{12\pi G \rho_0}
(t{-}t_0) \right]} \right\} \,, \label{Sc61}
\end{equation}
it describes the super-exponential expansion.
The Hubble function is monotonic
\begin{equation}
H(t) {=} \sqrt{\frac{8\pi G \rho(1)}{3}} \cosh{\left[\sqrt{12\pi G
\rho_0} \ (t{-}t_0) \right]} \label{Sc67}
\end{equation}
and increases infinitely.
The function $\dot{H}$
\begin{equation}
\dot{H}(t) {=} 4\pi G  \sqrt{2 \rho_0 \rho(1)} \sinh{\left[\sqrt{12\pi G
\rho_0} \ (t{-}t_0) \right]} \label{Sc679}
\end{equation}
also tends to infinity at $t \to \infty$.
The acceleration parameter
\begin{equation}
-q(t) = 1 {+} \sqrt{\frac{9\rho_0}{2\rho(1)}} \
\frac{\sinh{\left[\sqrt{12\pi G \rho_0} \ (t{-}t_0)
\right]}}{\cosh^2{\left[\sqrt{12\pi G \rho_0} \ (t{-}t_0)
\right]}} \label{Sc69}
\end{equation}
starts with ${-}q(t_0){=}1$, reaches the maximum ${-}q_{({\rm
max})}{=}1 {+} \sqrt{\frac{9\rho_0}{8\rho(1)}}$ at $t{=}t_0 {+}
\frac{\log{(1+\sqrt2)}}{\sqrt{12\pi G \rho_0}}$, and tends
asymptotically to ${-}q(\infty){=}1$. This is the explicit example of
the Little Rip.

\subsubsection{Complex roots [$(\sigma-3\xi)^2 < 12\xi$]}

Complex roots appear at positive $\xi$. The solution is quasiperiodic
$$
Z(x) = x^{-\gamma} \left\{Z(1)\cos{(\beta \log{x})} {+}
\right.
$$
\begin{equation}
\left.
{+}\frac{\left[Z^{\prime}(1){+} \gamma Z(1)  \right]}{\beta} \sin{(\beta \log{x})} \right\}  \,,
\label{period2}
\end{equation}
where
\begin{equation}
\gamma \equiv \frac{\sigma + 3
\xi}{2\xi} \,, \quad \beta \equiv \frac{1}{2\xi} \sqrt{12 \xi -
(\sigma -3\xi)^2} \,. \label{period1}
\end{equation}
In this section it is convenient to use the variable $\tau{=}\log{x}$ and to present the square of the Hubble function as follows
\begin{equation}
H^2(\tau) {=} \frac{8\pi G \rho_0}{3(1{+}\sigma)}\left\{
1{+} e^{{-}\gamma \tau}\left[{\cal A} \cos{\beta \tau}{+} {\cal B} \sin{\beta \tau} \right] \right\}
\,, \label{period3}
\end{equation}
where the parameters
\begin{equation}
{\cal A} = \sigma Z(1)+ \xi Z^{\prime}(1) \,, \label{period4}
\end{equation}
\begin{equation}
{\cal B} = \frac{1}{\beta}(\sigma -\gamma \xi)\left[Z^{\prime}(1){+} \gamma Z(1) \right] - \xi \beta Z(1)  \label{period5}
\end{equation}
depend on the initial data $Z(1)$ and $Z^{\prime}(1)$; for the illustration one can choose them so that ${\cal B}{=}0$.
Again there are three interesting cases.

\vspace{3mm}
\noindent
{\it (i) $H^2(\tau)>0$ for arbitrary $\tau$.}

\noindent
For instance, when $\gamma>0$, ${\cal B}{=}0$ and $|{\cal A}|<1$ we obtain that $H^2(\tau)>0$. Then the Hubble function $H(\tau)$ is real and tends to $H_{\infty} {=} \sqrt{\frac{8\pi G \rho_0}{3(1{+}\sigma)}}$ at $\tau \to \infty$; we deal with the Pseudo Rip.

\vspace{3mm}
\noindent
{\it (ii) $H^2(\tau)\geq 0$.}

\noindent
To illustrate the situation, for which $H^2$ is nonnegative and has a number of zeros, let us choose, first, that $\gamma{=}0$, second, that ${\cal A}{=}1$, ${\cal B}{=}0$, providing
\begin{equation}
H^2(\tau) {=} \frac{16\pi G \rho_0}{3(1{+}\sigma)} \cos^2{\frac{1}{2}\beta \tau} \geq 0
\,. \label{period6}
\end{equation}
The Hubble function is periodic with maximal value $H_{({\rm max})}{=}\sqrt{\frac{16\pi G \rho_0}{3(1{+}\sigma)}}$.
The corresponding scale factor
\begin{equation}
a(t) {=} a(t_0) \exp\left\{
\frac{2}{\beta} {\rm arcsin}\left[\tanh{\beta \sqrt{\frac{4\pi G \rho_0}{3(1{+}\sigma)}}(t{-}t_0)} \right]
\right\}
\label{period7}
\end{equation}
takes finite values satisfying the inequalities
\begin{equation}
e^{{-}\frac{\pi}{\beta}} \leq  \frac{a(t)}{a(t_0)} \leq e^{\frac{\pi}{\beta}}
\,. \label{period8}
\end{equation}
Thus, there are neither Big Rip, nor Little Rip, despite ${-}1<\sigma<0$ ($\xi<\frac{1}{3}$).

\vspace{3mm}
\noindent
{\it (iii) $H^2$ changes the sign.}

\noindent
When $\gamma <0$, the quasiperiodic function $H^2(\tau)$ given by (\ref{period3}) inevitably reaches zero value at some moment $\tau{=}\tau_{*}$.
Near this point the function $H^2(\tau)$ can be presented as
\begin{equation}
H^{2}(\tau) \simeq h(\tau_{*}{-}\tau) {+} \frac{1}{2}\omega (\tau_{*}{-}\tau)^2 {+} ...  \label{period9}
\end{equation}
Depending on the guiding parameters $\sigma$, $\xi$, $\rho_0$, and on the initial data $Z(1)$ and $Z^{\prime}(1)$,
the quantity $h$ can vanish or be positive.

\noindent
(1) $h \neq 0$.

\noindent
In the vicinity of the moment $\tau_{*}$ one obtains that $\tau_{*}{-}\tau {=} \frac{h}{4}(t{-}t^{*})^2$, and thus $a(t) \propto \exp{\left[{-}\frac{h}{4}(t{-}t^{*})^2 \right]}$.

\noindent
(2) $h{=}0$.

\noindent
The decomposition (\ref{period9}) starts with $(\tau_{*}{-}\tau)^2$, we obtain that
$a(t) \propto \exp{\left[1{-}e^{{-}\sqrt{\frac{\omega}{2}}(t{-}t_{*})} \right]}$.

In both cases the Universe has finite size and its life-time is finite.

\subsubsection{Resume}

To summarize the results of the analysis let us indicate the domains on the plane of the parameters $\xi$ and $\sigma$, for which the Big Rip scenaria can be realized: these domains are displayed on Fig.1.  First, let us focus on the case $\xi{=}0$. According to the results of Subsection IIA, the Big Rip can be realized on the interval ${-}1<\sigma<0$ of the vertical line $\xi{=}0$. It is the "classical" Big Rip domain, which is characterized by the condition $\rho{+}\Pi{=}\frac{1{+}\sigma}{\sigma}\rho <0$. When $0<\xi<\frac{1}{3}$ this "classical" Big Rip zone contracts along the line $\sigma$ to the interval ${-}1<\sigma<3\xi {-}2\sqrt{3\xi}$, and disappears at all when $\xi \geq \frac{1}{3}$ (see the domain I on Fig.1). In this sense, the simplest rheological property of the DE, namely, the retardation of the DE response to the Universe expansion, can avoid the Big Rip, which was the fate of the Universe at ${-}1<\sigma<0$, if the relaxation parameter $\xi$ exceeds the critical value $\xi{=}\frac{1}{3}$. In other words, when the DE relaxation time $\tau(t)\equiv \frac{\xi}{H(t)}$ is bigger than $\frac{1}{3H}{=}\frac{1}{\Theta}$, where $\Theta \equiv \nabla_k U^k$ is the Universe expansion scalar, the regime of the Big Rip can not be supported by the DE with such relaxation time.  If $\xi$ is negative, the interval ${-}1<\sigma<0$ again relates to the Big Rip scenario of the Universe expansion (see the domain II on Fig.1). The choice of the parameter $\rho_0$ can not avoid the Big Rip scenario, but it predetermined the type of the Big Rip: the power-law type, hyperbolic or trigonometric ones.

When $\sigma>0$, there was no classical Big Rip at $\xi{=}0$. The same fact can be indicated at $\xi>0$. However, if the parameter $\xi$ is negative, the Big Rip scenario is possible for $\sigma>0$, since instead of retardation the DE displays the acceleration of the response to the Universe expansion (see the domain III on Fig.1).

When $\sigma<{-}1$, there was no "classical" Big Rip at $\xi{=}0$. However, the Big Rip  becomes possible at $\sigma<{-}1$ for positive relaxation parameter (see the domain IV on Fig.1). Mathematically, this new domain for the Big Rip scenaria appears, since the characteristic equation is quadratic, and the presence of the second root in (\ref{roots17}) extends the possibilities of the Big Rip scenaria. Indeed, when $\xi \neq 0$, but $\xi \to 0$, the roots of the mentioned characteristic equation can be estimated as
\begin{equation}
s_1 \to -3 \frac{(1+\sigma)}{\sigma} \equiv \frac{1}{\alpha}\,, \quad s_2 \to - \frac{\sigma}{\xi} \,.\label{rootsqq1}
\end{equation}
The first root does not depend on $\xi$ and gives us the "classical" Big Rip, when ${-}1<\sigma<0$. The second root is positive and thus describes the "non-classical" Big Rip, when either $\sigma<0$ and $\xi>0$, or $\sigma>0$ and $\xi<0$. This explains the appearance of two new domains IV and V on Fig.1. For small values of $\xi$ the second characteristic number $s_2 \to {-} \frac{\sigma}{\xi}$ is respectively big; from the physical point of view, we can speak about instability of the DE response to the Universe expansion near the bifurcation line $\xi{=}0$.

{\bf
\begin{figure}
[htmb]
\includegraphics[width=8.25cm,height=8.501cm]{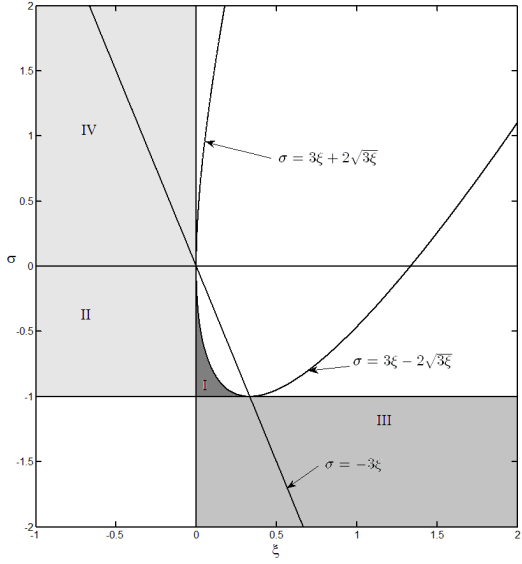}
\caption {
{\small The domains on the plane of the parameters $\xi$ and $\sigma$, for which the Universe evolution follows the Big Rip scenario in the model of the asymptotic dark energy domination.
The horizontal stripe ${-}1<\sigma<0$ is divided into three domains. There are two domains of the Big Rip in this stripe: the first domain is bounded by the part of parabola $\sigma{=}3\xi{-}2\sqrt{3\xi}$ ($0 \leq \xi<\frac{1}{3}$), by the vertical line $\xi{=}0$ and horizontal line $\sigma{=}{-1}$; the second domain is situated at $\xi<0$.
Two rectangular sectors $\xi>0$, $\sigma<{-}1$ and $\xi<0$, $\sigma>0$ also refer to the case of Big Rip.}
}
\end{figure}
}

\section{Big Rip avoidance at $ \nu_{*}\neq 0$, $\xi{=}0$}

In this model we suggest that the dark energy is described by the
equation of state $\rho = \rho_0 {+} \sigma \Pi$, but now the
Archimedean-type coupling is switched on. In other words, there
are two coupled  energy reservoirs in the Universe: the dark energy and
dark matter, and an effective energy redistribution between them
is possible in the course of the Universe accelerated expansion.

\subsection{Asymptotic behavior at $\Pi \to - \infty$}

Let the moment $t_0$ (or equivalently, $x{=}1$) be chosen so that
the value $\Pi(1)$ is large and negative. The term ${\cal J}$ (see
(\ref{key2})) takes the form
\begin{equation}
{\cal J} \to \frac{E_{*} \nu_{*} I_{*}}{x^3} \ e^{\nu_{*} \Pi(1)}
\ \Pi^{\prime}(x) \ e^{-\nu_{*} \Pi(x)} \,, \label{Ar1}
\end{equation}
where we use the new parameter
\begin{equation}
I_{*} \equiv \int^{\infty}_0 q^3 dq e^{{-}\lambda_{*} \sqrt{1{+}q^2}} {=}
\frac{2}{\lambda_{*}^2}
e^{{-}\lambda_{*}}\left(1{+}\frac{3}{\lambda_{*}}{+}\frac{3}{\lambda_{*}^2} \right)
\,. \label{Ar2}
\end{equation}
Here and below the symbols with asterisk: $I_{*}$, $E_{*}$, $\nu_{*}$, relate to the parameters of
the leading DM component. The key equation (\ref{key1}) transforms into
nonlinear differential equation of the first order
\begin{equation}
x \Pi^{\prime} \left[\frac{E_{*} \nu_{*} I_{*}}{x^{4}}
e^{\nu_{*} [\Pi(1)-\Pi]} {-}\sigma \right] {=} 3\left[\rho_0 {+} (1{+}\sigma) \Pi
\right]\,, \label{Ar3}
\end{equation}
and its structure prompts us the following transformations.

\subsubsection{New dynamic variables}

Let us introduce new variables $X$, $Y$ and parameter
$a_{*}$. For the case $\sigma>{-1}$ the transformations have the form:
\begin{equation}
X \equiv \frac{x^4}{a^4_{*}} \,, \quad Y \equiv - \nu_{*}
\left[\Pi + \frac{\rho_0}{1{+}\sigma} \right] \,,
 \label{Ar4}
\end{equation}
\begin{equation}
a^4_{*} \equiv
\frac{4E_{*} \nu_{*} I_{*}}{3(1{+}\sigma)} e^{-Y(1)} \,.
 \label{Ar44}
\end{equation}
When $\sigma<{-1}$, we replace formally $1{+}\sigma$ by $|1{+}\sigma|$ and $X$ by ${-}X$.
For the case $\sigma {=} {-}1$ the mentioned transformations are not appropriate and we consider this submodel in the special section.
In terms of new variables the key equation can be transformed into the linear
equation for the function $X(Y)$:
\begin{equation}
 Y \frac{d X}{d Y} = 4 \alpha X + e^Y \,.
 \label{Ar5}
\end{equation}
The energy density of the DE and DM in these new terms can be
written as
\begin{equation}
\rho = \frac{\rho_0}{1+\sigma} -\frac{\sigma}{\nu_{*}} Y \,, \quad
E = - \frac{\sigma}{4\alpha \nu_{*} X} e^Y \,, \label{Ar11}
\end{equation}
thus, the square of the Hubble function takes the form
\begin{equation}
H^2  = \frac{8 \pi G}{3}\left[\frac{\rho_0}{1+\sigma} -
\frac{\sigma}{\nu_{*}}\left(Y + \frac{1}{4\alpha X} e^Y \right)
\right] \,, \label{Ar12}
\end{equation}
and the rate of the Hubble function evolution $\dot{H}$ is described by the formula
\begin{equation}
\dot{H}  = \frac{4 \pi G (1{+}\sigma)}{\nu_{*}}\left[Y - \frac{1}{X} e^Y \right] \,. \label{Ar121}
\end{equation}
The function $a(t)$ can be extracted from the equation
\begin{equation}
\sqrt{\frac{128 \pi G}{3}}(t{-}t_0)  {=}  \int^{Y(t)}_{Y(1)}
\frac{dY \left[4\alpha {+} \frac{e^Y}{X(Y)} \right]}{Y
\sqrt{\frac{\rho_0}{1{+}\sigma} {-} \frac{\sigma}{\nu_{*}}\left(Y {+}
\frac{1}{4\alpha X} e^Y \right)}} \,. \label{Ar13}
\end{equation}
Here we have to put $X(Y)$ as the solution of (\ref{Ar5}) and to use
the inverse function
\begin{equation}
Y(t) \equiv Y[X(t)] = Y\left[\left(\frac{a(t)}{a(t_0)a_{*}}
\right)^{4} \right] \label{Ar14}
\end{equation}
on the upper limit of the integral.

\subsubsection{Qualitative analysis: divergence of the basic integral}

The solution to the equation (\ref{Ar5}) is
\begin{equation}
X(Y) = X(1) \left[\frac{Y}{Y(1)}\right]^{4\alpha} + Y^{4\alpha}
\int^Y_{Y(1)} \frac{du}{u^{1+4\alpha}} e^{u} \,. \label{Ar6}
\end{equation}
Here the constants $X(1)$ and $Y(1)$ are defined using the initial
data as follows:
\begin{equation}
X(1) = \frac{1}{a^4_{*}} \,, \quad Y(1) = - \nu_{*} \left[\Pi(1) +
\frac{\rho_0}{1+\sigma} \right] \,. \label{Ar70}
\end{equation}
Whatever the parameter $\alpha$ is, when $\Pi \to {-} \infty$
(i.e., $Y \to {+} \infty$) the leading order terms in the
asymptotic decomposition of the function $X(Y)$
\begin{equation}
X(Y) \to \frac{1}{Y} e^Y \left[1 {+} \frac{1{+}4\alpha}{Y} {+}
\frac{(1{+}4\alpha)(2{+}4\alpha)}{Y^2} {+} ... \right] \label{Ar7}
\end{equation}
come from the generalized integral exponential
\begin{equation}
E^{(\alpha)}(Y) \equiv  \int^Y_{Y(1)} \frac{du}{u^{1+4\alpha}}
e^{u} \,, \label{Ar8}
\end{equation}
which appears in the right-hand side of (\ref{Ar6}).
Thus, when $X \to \frac{1}{Y}e^Y$, the integrand in (\ref{Ar13}) behaves at
$Y \to \infty$ as
\begin{equation}
\sqrt{\frac{4\nu_{*}}{(3{-}\sigma)}} \ \frac{dY}{\sqrt{Y}} \,. \label{Ar15}
\end{equation}
This asymptotic estimation is valid when $\sigma<3$, i.e., in both interesting cases: ${-}1<\sigma<0$ and $\sigma<{-}1$.
Big Rip can exist, when the integral in the right-hand side of
(\ref{Ar13}) converges at $Y \to \infty$. In our case the integral
diverges at $Y \to \infty$ as $\sqrt{Y}$, and Big Rip can not be
realized in the  scenario with $\xi {=}0$ and $\nu_{*} \neq 0$.
In the asymptotic limit $Y \to \infty$ the formula (\ref{Ar12}) for the $H^2$ yields
\begin{equation}
H^2  \to  \frac{8 \pi G}{3}\left[\frac{\rho_0}{1+\sigma} + \frac{\sigma -3}{4\nu^*} +
\frac{(3-\sigma)}{4\nu_{*}}Y \right] \,. \label{aAr12}
\end{equation}
In other words, when $\sigma<3$ the Hubble function infinitely grows as $H \to \sqrt{\frac{(3-\sigma)}{4\nu_{*}}Y}$.
As for $\dot{H}$, (\ref{Ar121}) yields
\begin{equation}
\dot{H}  \to \frac{4 \pi G (3{-}\sigma)}{3\nu_{*}}\left[1+ \frac{1}{Y} \right] \,, \label{aAr121}
\end{equation}
i.e., the rate of the Hubble function growth tends to the positive constant $\dot{H}_{\infty}{=} \frac{4 \pi G (3{-}\sigma)}{3\nu_{*}}$.

\subsubsection{Search for the scale factor}

Keeping in mind (\ref{Ar13}) and (\ref{Ar15}) we can reconstruct now the function $a(t)$ using the function
\begin{equation}
Y(t) \to \frac{8 \pi G (3-\sigma)}{3 \nu_{*}}
(t-t_0)^2 \,, \label{Ar16}
\end{equation}
for the case $0 \leq \alpha < \infty$.
When $Y \to \infty$, the behavior of the function $X(Y)$ is
predetermined by the second term in (\ref{Ar6}), since the
generalized integral exponent (\ref{Ar8}) gives the exponential
leading order term $X \propto \frac{1}{Y}e^Y$ (see (\ref{Ar7})). The inverse function $Y(X)$
is of the logarithmic type, which can be presented by the following
iteration procedure
\begin{equation}
Y \to \log {XY} \to \log{[X
\log[X \log[X...]]]} \,, \label{Ar9}
\end{equation}
zero-order estimations yielding
\begin{equation}
Y \to \log{X Y(1)} \,. \label{Ar10}
\end{equation}
The  scale factor
\begin{equation}
a(t) \to \frac{a(t_0) a_{*}}{Y^{\frac{1}{4}}(1)} \exp\left\{\frac{2\pi G (3{-}\sigma)}{3\nu_{*}}
(t{-}t_0)^2 \right\}  \label{Ar17}
\end{equation}
is described by the anti-Gaussian function, which was obtained in \cite{Arch1}
as the exact solution of master equations of the Archimedean-type model.
Since the corresponding Hubble function
\begin{equation}
H(t) \to \frac{4\pi G (3{-}\sigma)}{3\nu_{*}} (t{-}t_0) \label{Ar199}
\end{equation}
grows linearly with time, the obtained solution can be classified as the Little Rip with $H\to \infty$ and $\dot{H}\to const$.

To complete the analysis let us consider two special cases.

\vspace{3mm}
\noindent
{\it (i) Special case $\sigma {=}{-}1$}

\noindent
In the special case: $\alpha{=}\infty$, or equivalently, $\sigma {=}{-}1$, we should consider the solutions in more details.
The key equation
\begin{equation}
x \Pi^{\prime} \left[\frac{E_{*} \nu_{*} I_{*}}{x^{4}}
e^{\nu_{*} [\Pi(1)-\Pi]} {+} 1 \right] {=} 3 \rho_0 \label{3Ar3}
\end{equation}
can be now transformed into
\begin{equation}
\frac{d\tilde{X}}{d\tilde{Y}}  = - \frac{4}{3\rho_0 \nu_{*}} \left(\tilde{X} + e^{\tilde{Y}} \right)
\,, \label{3Ar4}
\end{equation}
using the modified replacements
\begin{equation}
\tilde{X} \equiv \frac{x^4}{a^4_{**}} \,, \quad a^4_{**} \equiv E_{*}
\nu_{*} I_{*}  \,,
 \label{3Ar49}
\end{equation}
\begin{equation}
\tilde{Y} \equiv - \nu_{*} \left[\Pi - \Pi(1) \right] \,, \quad \tilde{Y}(1) = 0
\,.
 \label{3Ar44}
\end{equation}
The solution to (\ref{3Ar4})
\begin{equation}
\tilde{X}(\tilde{Y})= \left[\tilde{X}(1) {+} \frac{1}{1{+}\frac{3\rho_0 \nu_{*}}{4}}
\right] e^{{-}\frac{4}{3\rho_0 \nu_{*}} \tilde{Y}} {-}
\frac{1}{1{+}\frac{3\rho_0 \nu_{*}}{4}} e^{\tilde{Y}} \,,
 \label{3Ar45}
\end{equation}
shows that $\tilde{X}(\tilde{Y})$ takes zero value at some $\tilde{Y} {=} \tilde{Y}_{*}$ and then
changes the sign. Since $\tilde{X}$ is a positively defined function, we
can conclude that the regime $a \to \infty $ with $\Pi \to -
\infty$ can not be realized, when $\sigma {=}{-}1$. In other
words, the phantom-crossing value $\sigma{=}{-}1$, which
corresponds to $\alpha{=}\infty$ is the critical value: when
${-}1< \sigma\leq 0$, the Little Rip regime with $a(t) \to \infty$
and $\Pi(t) \to {-} \infty$ is possible; when $\sigma {=}{-}1$,
the scale factor can not reach infinite value.

\vspace{3mm}
\noindent
{\it (ii) Special case $\sigma{=}3$}

\noindent
The special case: $\alpha{=}{-}\frac{1}{4}$, or equivalently, $\sigma {=}3$, relates to the ultrarelativistic DE, when $\rho_0{=}0$, since $\rho{=}3\Pi$.
The functions $X$, $H^2$ and $\dot{H}$ take now the form
\begin{equation}
X = \frac{1}{Y}\left(K + e^{Y} \right) \,,
 \label{333}
\end{equation}
\begin{equation}
H^2 = \frac{2\pi G}{3}\left[\rho_0 - \frac{12 K Y}{\nu_{*}(K+e{Y})}\right]
\,,
 \label{3332}
\end{equation}
\begin{equation}
\dot{H} = \frac{16 \pi G K}{\nu_{*}\left(K + e^{Y} \right)} \ Y
\,,
 \label{3333}
\end{equation}
where the new constant $K$ is introduced
\begin{equation}
K \equiv X(1)Y(1) - e^{Y(1)}
\,.
 \label{3339}
\end{equation}
When $Y\to \infty$, $H \to \sqrt{\frac{2\pi G \rho_0}{3}}$, $\dot{H} \to 0$, thus, we deal with the de Sitter asymptote.

\section{Big Rip avoidance in general case: $\xi \neq 0$ and $\ \nu_{*} \neq 0$}

In terms of the dynamic variables (\ref{Ar4}) the key equation for the DE pressure takes the form
\begin{equation}
\frac{16 \xi}{3(1{+}\sigma)} X^2 Y^{\prime \prime} {+} Y^{\prime}
\left[\frac{4(\sigma{+}7\xi)}{3(1{+}\sigma)} \ X {-} e^Y \right] {+} Y =0
\,.
 \label{5Ar1}
\end{equation}
At $\xi {=}0$ it coincides with  (\ref{Ar5}), as it should be.
The function $\dot{H}$ and the square of the Hubble function can be presented as follows
\begin{equation}
\dot{H} = \frac{4\pi G}{\nu_{*}} \left[(1+\sigma) \left(Y-\frac{e^{Y}}{X} \right)+ 4\xi X \frac{dY}{dX}
\right]
\,,
 \label{es1}
\end{equation}
\begin{equation}
H^2  = \frac{8 \pi G}{3}\left[\frac{\rho_0}{1{+}\sigma} {-}
\frac{\sigma}{\nu_{*}}\left(Y {+} \frac{1}{4\alpha X} e^Y \right)
{-} \frac{4\xi}{\nu_{*}} X \frac{dY}{dX}
\right]
\,.
 \label{es2}
\end{equation}
Let us, first, analyze these functions in terms of the variable $Y$.

\subsection{Qualitative analysis}

\subsubsection{The model with $0<\xi<\frac{1}{3}$ and ${-}1<\sigma <0$}

Let us assume that the asymptotic solution of (\ref{5Ar1}) at $Y \to
\infty$ differs insignificantly from the solution to (\ref{Ar5}), and
let us consider the function
\begin{equation}
X(Y) \to \frac{1}{Y} e^Y \left[1+ \frac{B(\xi)}{Y} \right] \,,
\label{5Ar21}
\end{equation}
in which the parameter $B(\xi)$ is unknown, but should satisfy the
condition
\begin{equation}
B(0) = 1+ 4\alpha \,.
 \label{5Ar2}
\end{equation}
This means that we consider the solution in the form (\ref{Ar7})
and restrict ourselves by the term $\frac{1}{Y}$ in the
parentheses. With mentioned accuracy the function $X(Y)$
(\ref{5Ar21}) satisfies (\ref{5Ar1}) with (\ref{5Ar2}) , when
\begin{equation}
B(\xi) = 1 - \frac{4(\sigma + 3\xi)}{3(1+\sigma)} \,.
 \label{5Ar29}
\end{equation}
From the qualitative point of view the behavior of the
DE with $\xi \neq 0$ at $\Pi \to {-}\infty$ is analogous to that of the DE
with $\xi{=}0$ and effective parameter $\sigma^{*}$ equal to
\begin{equation}
\sigma^{*} \equiv \frac{\sigma + 3\xi}{1-3\xi} \,.
 \label{5Ar296}
\end{equation}
It is interesting that $\sigma^{*}{=}{-}1$, when $\sigma{=}{-}1$,
i.e., these two parameters coincide for arbitrary $\xi$ at the
phantom-crossing point $\sigma {=}\sigma^{*}{=}{-}1$.

For the solution of the type (\ref{5Ar21}) the term $X\frac{dY}{dX}$ behaves as $(1{+}\frac{1}{Y})$, when $Y \to \infty$. Thus,
the contributions of the last terms in (\ref{es1}) and (\ref{es2}) (proportional to the parameter $\xi$) can be neglected at $Y \to \infty$.
Thus, the main conclusion that the Archimedean-type interaction avoids the Big Rip remains valid at $0<\xi<\frac{1}{3}$, ${-}1<\sigma <0$.

\subsubsection{The model with $\xi>0$ and $\sigma <{-}1$}

The key equation for this case can be obtained from  (\ref{5Ar1}) by the formal replacement $X \to {-}\tilde{X}$ and $Y^{\prime}(X) \to {-}Y^{\prime}(\tilde{X})$
with positively defined $\tilde{X}$. Numerical calculations show that in this case the regime $Y \to \infty$ does not exist.
To illustrate qualitatively such a behavior one can mention the following:  the regime with $Y \to \infty$ would require that
the leading order terms in the equation (\ref{5Ar1}) are linked by $ Y^{\prime} e^{Y}{+}Y {=}0$, which is in contradiction with the requirement $\tilde{X}>0$.
Then two scenaria are available: first, with $Y \to const$, second with $Y \to {-}\infty$. In the first scenario there is no Big Rip, and, as a rule, we deal with the Pseudo Rip. In the second scenario the quantity $Y$ with positive initial value $Y(1)>0$ would have negative derivative, thus, the equation
(\ref{es1}) would be transformed into
\begin{equation}
\dot{H} = -\frac{4\pi G}{\nu_{*}} \left[|1+\sigma|\left(Y+\frac{e^{Y}}{\tilde{X}} \right)+ 4\xi X \left|\frac{dY}{dX}\right|
\right]
\,,
 \label{es4}
\end{equation}
showing explicitly that the derivative of the Hubble function is negative. This means that at some moment $t^{*}$ (finite or infinite) the Hubble function will take zero value providing the scale factor $a(t)$ to reach its maximal value  $a_{{{\rm max}}}$. In other words, in this scenario the Big Rip can nor appear in contrast to the model with $\nu_{*}{=}0$.

\subsubsection{The model with $\xi<0$ and $\sigma >{-}1$}

When $\xi$ is negative, the procedure similar to the one used in the case $0<\xi<\frac{1}{3}$, ${-}1<\sigma <0$ gives the asymptotic formulas analogous to (\ref{aAr12}) and (\ref{aAr121})
\begin{equation}
\dot{H}(Y \to \infty) \to \frac{4\pi G}{3\nu_{*}} (3-\sigma)
\,,
 \label{es48}
\end{equation}
\begin{equation}
H^2(Y \to \infty ) \to \frac{2\pi G (3{-}\sigma)}{3\nu_{*}} \ Y
\,,
 \label{es5}
\end{equation}
which do not contain the parameter $\xi$ in the leading order term. However, now we consider the parameter $1{+}\sigma$ to be positive, and can conclude that we deal with the Little Rip, when ${-}1<\sigma<3$, and the solution with finite $|a(t)|$, when $\sigma>3$. In any case the Archimedean-type interaction avoids the Big Rip.

\subsubsection{The special model $\sigma {=}{-}1$}

As in the general case described in \cite{Arch1,Arch2}, asymptotic
model admits the solution of the anti-Gaussian type, when $\sigma
{=}{-}1$. Indeed, the corresponding key equation
\begin{equation}
\xi x^2 \Pi^{\prime \prime} + (4\xi-1) x \Pi^{\prime} + 3\rho_0 =
\frac{E_{*}\nu_{*}I_{*}}{x^3}
\Pi^{\prime}e^{-\nu_{*}\left[\Pi(x)-\Pi(1) \right]}
 \label{AG1}
\end{equation}
is satisfied with the solution
\begin{equation}
\Pi(x) = \Pi(1) - \frac{4}{\nu_{*}} \log{x} \,,
 \label{AG2}
\end{equation}
if the guiding parameters are coupled by the equality
\begin{equation}
 E_{*}I_{*} = \frac{3\xi -1}{\nu_{*}}- \frac{3}{4} \rho_0 \,.
 \label{AG3}
\end{equation}
Since $E_{*}I_{*}$ is positive, one requires that $\xi>
\frac{1}{3}+ \frac{1}{4}\rho_0 \nu_{*}$, i.e., this equality can
not be satisfied when $\xi{=}0$. When the right-hand side of
(\ref{AG3}) is positive, the logarithmic solution can appear as a
result of some "fine tuning": one can satisfy (\ref{AG3}), e.g.,
by varying the initial temperature $T_{*}$ of the DM, which enters the
parameter $\lambda_{*}{=}\frac{m_{*}c^2}{k_B T_{*}}$, the argument of the
function $I_{*}(\lambda_{*})$ (see (\ref{Ar2})). The asymptotic DE
energy-density is now
\begin{equation}
\rho(x) = \rho(1) + \frac{4}{\nu_{*}} \log{x} \,, \label{toy11}
\end{equation}
where the relation
\begin{equation}
\Pi^{\prime}(1) = \frac{1}{\xi}\left[ \rho(1) - \rho_0 + \Pi(1)
\right] \label{toy011}
\end{equation}
is used. The sum $\Pi(x) {+} \rho(x)$ remains constant
\begin{equation}
\rho(x) + \Pi(x) = \rho(1) + \Pi(1) = \rho_0 -
\frac{4\xi}{\nu_{*}} \,. \label{toy2}
\end{equation}
The DM energy density $E(x)$ is now constant, i.e.,
\begin{equation}
E(x) = E_{*}I_{*} \,, \label{toy3}
\end{equation}
and the Hubble function $H(x)$ can be found from the
equation
\begin{equation} H^2(x) = \frac{8\pi G}{3} \left[\rho(1)
+ E_{*}I_{*} + \frac{4}{\nu_{*}} \log{x} \right] \,. \label{toy4}
\end{equation}
The corresponding scale factor $a(t)$ can be written in the form
\begin{equation}
a(t) = a^{*} \exp\left\{\frac{8\pi G}{3 \nu_{*}} (t-t^{*})^2
\right\} \,, \label{toy5}
\end{equation}
where the parameters with asterisks are defined as follows
$$
a^{*} \equiv a(t_0) \exp\left\{- \frac{\nu_{*}}{4}[\rho(1)+
E_{(0)}] \right\}  \,,
$$
\begin{equation}
t^{*} \equiv t_0 - \nu_{*} \sqrt{\frac{3}{32 \pi G}[\rho(1)+
E_{(0)}]} \,. \label{toy6}
\end{equation}
The acceleration parameter $-q(t)$ for the anti-Gaussian expansion
is positive and exceeds the unity:
\begin{equation}
- q(t) \equiv \frac{\ddot{a}}{a H^2} = 1 + \frac{3 \nu_{*}}{ 16
\pi G (t{-}t^{*})^2}  \  \geq 1 \,. \label{toy7}
\end{equation}
Clearly, this solution illustrates one of the versions of the
Little Rip.

\subsubsection{Numerical analysis}

In order to confirm the qualitative conclusions made for the model with $\xi \neq 0$, $\nu_{*} \neq 0$, we analyzed numerically the models for the parameters $\xi$ and $\sigma$ belonging to the domains I,II,III,IV displayed on Fig.1. These calculations are illustrated by Fig.2 and Fig.3. These figures contain four panels: the first one displays the Hubble function $H(x)$, the second panel demonstrates the behavior of the scale factor $a(t)$, the third panel describes the rate of the growth of the Hubble function $\dot{H}$, the fourth panel displays the function $\rho(x)$.

{\bf
\begin{figure}
[htmb]
\includegraphics[width=8.25cm,height=9.627cm]{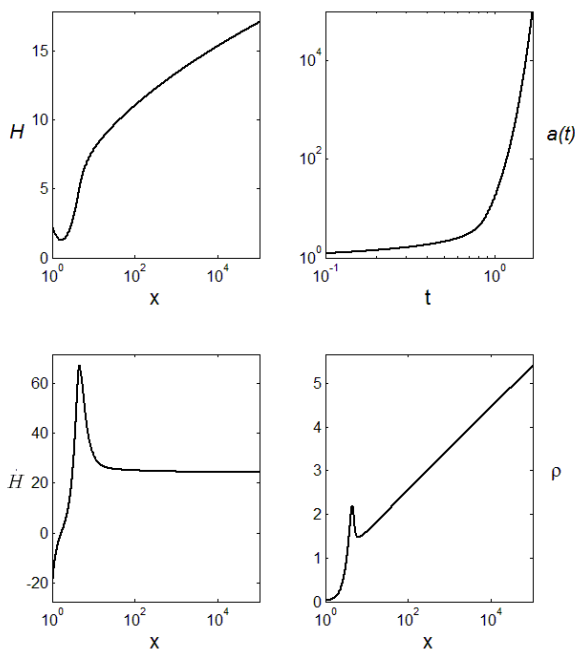}
\caption {
{\small The illustration to the model with $\sigma {=}{-}0.1$, $\xi {=}{-}0.1$, $\nu_{*}{=}1$ (i.e., for the case ${-}1<\sigma<0$, $\xi<0$ and $\sigma{+}3\xi<0$, which refers to the Big Rip scenario at $\nu_{*}{=}0$ according to Fig.1). Clearly, when $\nu_{*}\neq 0$, i.e., the Archimedean-type coupling is switched on, the model solution is of the Little Rip type with $a\to \infty$, $H \to \infty$, $\dot{H}\to const$, $\rho \to \infty$, $\Pi \to {-}\infty$.}
}
\end{figure}
}

{\bf
\begin{figure}
[htmb]
\includegraphics[width=8.25cm,height=9.627cm]{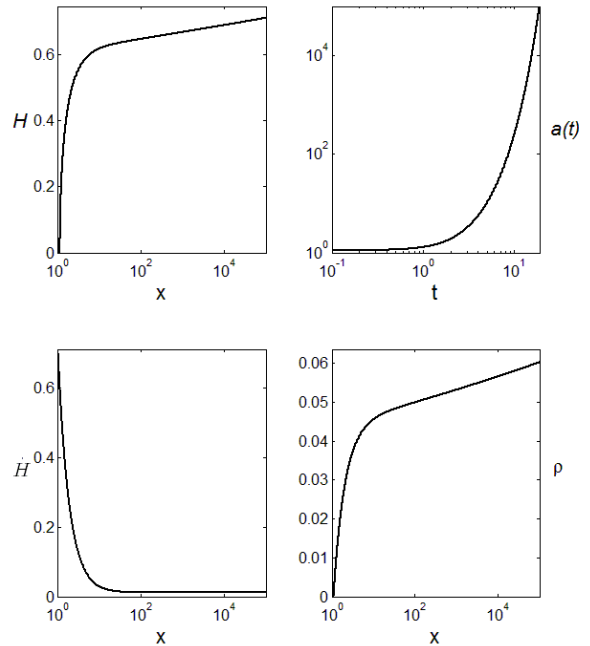}
\caption {
{\small The illustration to the model with $\sigma=-1.01$, $\xi=0.7$, $\nu_{*}{=}1$ (i.e., for the case $\sigma<{-}1$, $\xi>0$ and $\sigma{+}3\xi>0$, which refers to the Big Rip scenario at $\nu_{*}{=}0$ according to Fig.1). When $\nu_{*}\neq 0$ the model solution is of the Little Rip type with $a\to \infty$, $H \to \infty$, $\dot{H}\to const$, $\rho \to \infty$, $\Pi \to {-}\infty$.}
}
\end{figure}
}

\section{Discussion}

When the dark matter is effectively coupled to the dark energy by the Archimedean-type interaction, the late-time evolution of the Universe with negative DE pressure is protected from the Big Rip singularity, and the Little Rip becomes a typical fate of the Universe. This is the main conclusion of the paper. More detailed discussion includes the following three items.

\noindent
1. In the presented model of Archimedean-type coupling between the dark matter and dark energy the DE pressure, $\Pi$, is the key element of modeling. All the model solutions can be divided into
three classes with respect to asymptotic behavior of the state function $\Pi(t \to t_{\infty}){=} \Pi_{\infty}$ (we consider both cases: $t_{\infty} {=}t_s$ (future finite time singularity)  and
$t_{\infty} {=} \infty$). The first class is characterized by finite $\Pi_{\infty}$; the second class relates to $\Pi_{\infty}{=}{+}\infty$; we deal with the solutions of the third class, when $\Pi_{\infty}{=}{-}\infty$. In \cite{Arch2,Arch3} we focused on the solutions of the first class, and have shown qualitatively that the solutions of the second and third classes exist for some values of the effective guiding model parameters $\xi$, $\sigma$, $\rho_0$ and $\nu_{*}$. Three parameters $\xi$, $\sigma$ and  $\rho_0$ describe the equation of state of the dark energy (see (\ref{simplest0})), $\nu_{*}$ is the effective Archimedean-type coupling constant. This paper is devoted to the analysis of the solutions of the second and third classes.
When $\Pi_{\infty}$ is finite or $\Pi_{\infty}{=}{+}\infty$, the dark matter is asymptotically decoupled from the dark energy and its energy density becomes negligible in comparison with the DE energy density. We indicated this asymptotic situation as the DE domination (the coupling constant $\nu_{*}$ is not equal to zero, but it becomes a {\it hidden} parameter of the model). In this case only the appropriate choice of the constitutive parameters $\xi$, $\sigma$ and  $\rho_0$ can protect the Universe from the Big Rip singularity. To summarize the analytical results for the DE domination epochs we prepared Fig.1.

When $\xi{=}0$, the Big Rip can be realized on the interval ${-}1<\sigma<0$ (or in other words $w{=}\frac{1}{\sigma}<{-1}$). It can be indicated as the "classical" Big Rip domain. When $0<\xi<\frac{1}{3}$ this "classical" Big Rip zone contracts along the line $\sigma$ to the interval ${-}1<\sigma<3\xi {-}2\sqrt{3\xi}$, and disappears at all when $\xi \geq \frac{1}{3}$. In this sense, the retardation of the DE response to the Universe expansion, can avoid the Big Rip, if the relaxation parameter $\xi$ exceeds the critical value $\xi{=}\frac{1}{3}$. In other words, when the DE relaxation time $\tau(t)\equiv \frac{\xi}{H(t)}$ is bigger than $\frac{1}{3H}{=}\frac{1}{\Theta}$, where $\Theta \equiv \nabla_k U^k$ is the Universe expansion scalar, the regime of the Big Rip can not be supported.  If $\xi$ is negative, the interval ${-}1<\sigma<0$ again refers to the Big Rip scenario of the Universe expansion. The choice of the parameter $\rho_0$ can not provide the avoidance of the Big Rip scenario, but it predetermined the type of the Big Rip: the power-law type, hyperbolic or trigonometric ones.
When $\sigma>0$, there was no "classical" Big Rip at $\xi{=}0$. The same fact can be indicated at $\xi>0$. However, if the parameter $\xi$ is negative, the Big Rip scenario is possible for $\sigma>0$, since, instead of retardation, the DE displays the acceleration of the response to the Universe expansion.
When $\sigma<{-}1$ (or, in other words, ${-1}<w<0$), there was no "classical" Big Rip at $\xi{=}0$, however, the Big Rip  becomes possible at $\sigma<{-}1$ for positive relaxation time parameter, $\xi>0$.

\noindent
2. The most interesting case studied in the paper concerns the model with asymptotic
behavior $\Pi(t \to t_{\infty}) \to {-}\infty$, for which the Archimedean-type coupling leads to the effective heating of the dark matter component of the dark fluid. Due to the force proportional to the four-gradient of the DE pressure the DM becomes effectively ultrarelativistic and thus plays an active role in the energy redistribution processes inside the dark fluid.
The reviving of this second player in the late-time scenario  of the Universe evolution change essentially the character of expansion: at $\nu_{*}\neq 0$ the Big Rip scenaria happen to be {\it avoided} and instead of  them the Little Rip scenaria become typical for the Universe late-time evolution. This avoidance is typical both for the cases $\xi{=}0$ and $\xi \neq 0$.

We would like to emphasize that the so-called anti-Gaussian type solutions for the scale factor $a(t)$ appearing as some specific exact solution in \cite{Arch1}, become typical asymptotic solutions in the case under discussion (see, e.g., (\ref{Ar17}) and (\ref{toy5})). These anti-Gaussian type solutions correspond to the Little Rip scenario \cite{LR1,LR5}, since $a(t\to\infty)\to \infty$,  $H(t\to\infty)\to \infty$, $|\Pi(t\to\infty)|\to \infty$ and $\rho(t\to\infty)\to \infty$. In other words, in this model the life-time of the Universe is infinite, the Hubble function and scale factor tend to infinity without vertical asymptotes.

Discussing the mechanism of the Big Rip avoidance in the presence of the Archimedean-type interaction, we would like to attract the attention to the following feature. When the coupling constant $\nu_{*}$ vanishes, the DE pressure behaves as the power-law function of the scale factor. In terms of dimensionless variables $Y$ and $X$ (see (\ref{Ar4})) this power-law function is predetermined by the first term in (\ref{Ar6}). When $\nu_{*} \neq 0$, and the DM contribution to the total energy-density gains the same order as the DE energy-density, the DE pressure behaves according to the super-logarithmic law (\ref{Ar9}), which is typical for the Little Rip. Super-logarithmic law for $Y$ means that the DE pressure grows more slowly than in the case of $\nu_{*}{=}0$. In other words, since the DE transfers energy to the DM due to the Archimedean-type interaction, the rate of the DE pressure growth decreases, thus protecting the Universe from the Big Rip singularity.

\vspace{3mm}
\noindent
3. The model under discussion displays that there are three {\it critical} values of the guiding parameters. The first one is the value $\sigma{=}{-}1$, which corresponds to the well-known phantom-crossing value $w{=}\frac{1}{\sigma}{=}{-}1$ for the DE equation of state parameter $w$. The symptom of criticality is that the term $(1{+}\sigma)^{{-}1}$ systematically appears in the key equations and expressions (see, e.g., (\ref{key1119}) and (\ref{key0167})). The manifestation of the criticality is that the solutions with $\sigma \to {-}1$ differ principally from the ones with $\sigma \equiv {-}1$, so that we considered the last case as the special one.
In this sense, one can see some analogy between this model and the critical behavior of the magnetic and electric fields affected by the gravitational wave: we obtained principally different solutions for the case, when the refractive index tends to one, $n \to 1$, and for the case, when this parameter is equal to one identically, $n \equiv 1$, since the term $n^2{-}1$ appeared in the denominators of the sought-for functions \cite{singular}.
The second critical value is $\nu_{*}{=}0$. Again, the parameter $\nu_{*}$ appears in the denominators in the functions $a(t)$ and $H(t)$ (see, e.g., (\ref{Ar17}), (\ref{Ar199}), as the symptom of criticality, and the solutions with $\nu_{*}{=}0$ (we mean the solutions of the Big Rip type) differ principally from the solutions with $\nu_{*}\neq 0$ (the solutions of the Little Rip type). The third critical value is connected with the parameter $\xi$. The value $\xi{=}0$ can be considered as the critical one, since crossing the vertical line $\xi {=}0$ on Fig.1, we see the structural rearrangement of the domains corresponding to the Big Rip type solutions. In addition, according to (\ref{rootsqq1}), this parameter appears in the denominator, indicating that the situation with $\xi \to 0$ (there are two roots of the characteristic equation) differs essentially from the situation with $\xi \equiv 0$ (there exists only one root).

The first application of the model was considered in \cite{Arch3}; we hope to present new results concerning the cosmological applications of the model in the next papers.

\begin{acknowledgments}
The authors are grateful to Prof. S.D. Odintsov for valuable comments.
This work was partially supported by the Federal Targeted Programme ``Scientific and Scientific - Pedagogical Personnel of the Innovative Russia''
(grants Nos 16.740.11. 0185 and  14.740.11. 0407), and by the Russian Foundation for Basic Research (grants Nos. 11-02-01162 and 11-05-97518 - p-center-a).
\end{acknowledgments}

\end{document}